\documentclass[doublecol]{epl2}

\usepackage{graphicx}
\usepackage{dcolumn}
\usepackage{bm}
\usepackage{subfigure}

\graphicspath{{figures/}}

\title{Equalitarian Societies are Economically Impossible}
\shorttitle{Equalitarian Societies are Economically Impossible}

\author{Bojin Zheng*\inst{1,2,3} \and Wenhua Du\inst{4} \and Wanneng Shu\inst{1}\and Jianmin Wang\inst{3} \and Deyi Li*\inst{3}
}

\institute{
  \inst{1} College of Computer Science,
South-Central University for Nationalities, Wuhan 430074, China \\
  \inst{2} State Key Laboratory of  Networking and Switching Technology, Beijing University of Posts and\\ Telecommunications, Beijing 100876, China\\
  \inst{3} School of Software, Tsinghua University, Beijing 100084, China\\
  \inst{4} Management School, South-Central University for Nationalities, Wuhan 430074, China
}


\abstract{
The inequality of wealth distribution is a universal phenomenon in the civilized nations, and it is often imputed to the Matthew effect, that is, the rich get richer and the poor get poorer. Some philosophers unjustified this phenomenon and tried to put the human civilization upon the evenness of wealth. Noticing the facts that 1) the emergence of the centralism is the starting point of human civilization, i.e., people in a society were organized hierarchically, 2) the inequality of wealth emerges simultaneously, this paper proposes a wealth distribution model based on the hidden tree structure from the viewpoint of complex network. This model considers the organized structure of people in a society as a hidden tree, and the cooperations among human beings as the transactions on the hidden tree, thereby explains the distribution of wealth. This model shows that the scale-free phenomenon of wealth
distribution can be produced by the cascade controlling of human society, that is, the inequality of wealth can parasitize in the social organizations, such that any actions in eliminating the unequal wealth distribution would lead to the destroy of social or economic structures, resulting in the collapse of the economic system, therefore, would fail in vain.
}

\begin{document}
\maketitle

The inequality of the wealth distribution is a universal phenomenon in the civilized nations, that is, most of people would have little wealth, and a few people would hold most. All the countries during most of the historical periods would obey this law\cite{203}, including the ancient Egyptian dynasty\cite{abulmagd-2002-66}, the modern Europe, the up-to-date USA, Japan, India\cite{chatterjee-2007-92}.

The inequality is often regarded to following the Matthew effect, which says that - the rich get richer and the poor get poorer. This effect implies that most of people, the poor, will face a more and more miserable circumstance; and the rich, will thrive forever. The Matthew effect claims an irresistible process and is a popular mechanism explanation to the inequality of wealth.

Numerous philosophers unjustified this inequality phenomenon and believed that all the creatures should be equally treated and be assigned the equal wealth. Under this idea, some nations were set up, for examples, the Soviet Union, the Taiping Heavenly Kingdom (1851-1864 in southern China) and other governments. These governments legislated the equal wealth distribution, but they failed finally due to serious economic depressions. Seemly the inequality phenomenon is natural and inevitable.

But why such an inequitable phenomenon can dominate human society? Does there exist a solution that can put the civilization upon the fair wealth distribution? Scientists have been addressing these problems.

Italian economist Vilfredo Pareto was a pioneer scientist. He proposed that the personal wealth would obey the power-law distribution\cite{29}. Then Yule proposed a mechanism named ``Yule Process'' to explain many scaling phenomena in various fields based on ``growth''\cite{28,30}. Obviously, the growth implies that the power-law distribution would depend on the lapse of time. Sorin Solomon et al. convinced that current personal wealth distribution is power-law\cite{205,206,207}, they also suggested that market is the source of the extreme wealth inequality, and used the Generalised Lotka Volterra model\cite{208} to explain this phenomenon. Furthermore, Solomon et al. explored the relationship between market size and market instability\cite{210}, and drew a conclusion that world-size global markets lead to economic instability\cite{209}.

In general, a lot of researches impute the origin of inequality to the market, or the investment. With the lapse of time, the market or the investment would form a positive feedback, which make the rich more successful in the future. Mathematically, this explanation can also fit the curve of wealth distribution quite well.

On the other hand, the researches on complex network can contribute some useful ideas to our understanding of the wealth distribution\cite{1,214,215}. All of us knew that the complex systems can often be modeled as Complex Networks\cite{168}. Since there exist many scaling phenomena in complex systems, hence, some complex networks, such as the Scientific Collaboration Networks\cite{31}, the World Wide Web\cite{1} etc., are scale-free. That is, their degree distributions obey the power-law distribution. To illustrate the mechanism of the Scientific Collaboration Networks, Price proposed the concept of accumulative advantage\cite{31}. Later, Merton coined the accumulative advantage processes as the Matthew effect\cite{96}. To explain the mechanism of the World Wide Web, the pioneer scientist Barab\'{a}si et al. also proposed a model named the growth with Preferential attachment. Both explanations are similar to the positive feedback.  For the scale-free networks, notice that when we map the degree to the fortune, so that the mechanisms of the scale-free networks can be regarded as the mechanism of wealth distribution and explain the origin of wealth inequality.


From above, the positive feedback, or the Matthew effect is a popular and reasonable explanation.

However, the Matthew effect is more likely a phenomenon, not the fundamental fact.

Seemly, when we focus on the phenomenon of wealth condensation,
this phenomenon is easy to understand, since the rich can invest in new sources of creating wealth, so that
the richer the rich are, the more they can earn. This process would repeat and repeat, the richer would get richer.


But the reality is not like this. John D. Rockefeller was the first billionaire on the earth, but his
heirs have to envy the eminence of their ancient now, even if they are smart enough and the economy of USA
increases continually. Not only Rockefeller family, the other richest families in that age also encounter
this problem. Of course, the richest men/women in current world can not yet escape from this principle. Probably, at least, the Matthew effect does not work on the richest families. The winners can not take all, the poor can become the richest, for example, the founders of Google, Facebook and Twitter etc..


On the other hand, suppose that we leave a small amount of money on deposit and then sleeping for several hundred
years, theoretically our offsprings will have a great amount of wealth through the action of compound interest.
Unfortunately, historical real rates of return show that the effects of taxation and inflation would likely
cut the money off, less than the amount that had been deposited. This result implies that the Matthew effect does not hold in
long run and does not work on the common population.


From the cases above, seemly the Matthew effect is often a theoretical process from the viewpoint of investment.
Considering that most of current top richers in USA didn't accumulate their fortune from the capital market,
this viewpoint may not be fully reasonable for all circumstances.

Actually, the inequality of wealth has an another source. The economists have deeply explored a universal phenomenon, the hierarchy in the firms\cite{216,217}. With the hierarchy, the company managers are paid with much more salaries and bonus than
their less paid employees, however, the owners of a company would have much more profit than the salaries
and bonuses of the managers. Here, the social position is positively related to the amount of wealth. Furthermore, if we regarded the society as a set of companies, then the wealth distribution could rely on the
hierarchy of company. That is, the disparity of wealth may come from the structure of company.

If the disparity of wealth could come from the structure of companies, strictly, the tree-like cascade controlling structure, that means we cannot eliminate this kind
of disparity, unless we destroy the structure. That is, if there is no structure, i.e., no company, so that no labor division, the society
would return to the prehistorical state, hence, economical products would fall into an ignorable level. That is, the hypothesis on cascade controlling will not only explain the wealth inequality, but also tell the reason that the inequality is natural and infrangible.


In this paper, we design a model based on the hidden tree structure\cite{171,151}, and proved that the trading actions on this structure can
produce a power law distribution of wealth, meaning that the social cascading controlling structure is one origin of wealth inequality. In contrast to the explanation to the Matthew effect based on
investment, the hidden structure model can explain more phenomena on wealth distribution, and also produces phenomenon similar to the Matthew effect.

\section{Modeling}
When we consider the emergence of the wealth, we can find that the wealth obviously relates to the structure in
ancient society. For example, the cacique in a tribe would probably be the richest man in the eolithic age, and
the biggest slave owner, the king would probably be the richest man in slavery society. From the
experiences of early human societies, the wealth may be closely positively related to the social
position in the society. In modern societies, the richest men actually are the kings in their fields.


Why does the wealth strongly relate to the social position? We can use a simple example to illustrate it. If
one find that he can buy some apples from the providers and then sell them to customers to make money, he
would repeat his business, if the profit is enough. When the market is big enough, he would find that he can
employ another person to help him. Obviously, the profit which is made by the employer would be shared
between the employee and the employer. The larger the business is, the more the employees can be employed, until
the employer can employ managers to help him, so a tree structure is built. In this process, the wealth
strongly relate to the social position, does not strictly relate to the initial wealth distribution, because
of the cascade controlling.  In this case, the poor does not have a poor destiny, and the rich does not have
 a fate to be richer. We can find many similar cases in the real world. In a sense, this kind of tree-like cascade controlling, actually is a hidden order\cite{43,151,171} in the society, therefore, we termed it as ``hidden tree structure''.

 The hidden tree structure means that: 1) it is a hierarchical structure and reflects the social positions of the
 individuals; 2) it is cascade controlling and by it the random actions are organized; 3) it underlies in the social phenomena and is the hidden order.


The hidden tree structure can produce the phenomenon of the Matthew effect. When the business of employer is growing bigger and bigger, and therefore richer and richer in contrast of
his employees, even if the employer does not have a lot of cashes in the whole process. Of course, the employer would hold much more fortune than the employees finally.

Though the effect of this mechanism satisfies the Matthew effect roughly, whether this mechanism can explain the
wealth distribution? If it works, the distribution made by it should obey the power law distribution.



Assume that the society is an isolated system, every person in this system is organized into a tree
structure. In the society, everyone survives on the services of the others. That means, everyone should trade
with the others. When two persons have a transaction, they have to take orders from their employers. To minimize the
cost, the participants should be minimized. That is, both sides in a transaction would choose the shortest route. Because of cascade controlling, this shortest path actually is the path between these two persons in
the tree structure.

Therefore, we design a model as follows.

\begin{figure}[htbp]
\includegraphics[width=8cm,height=4cm]{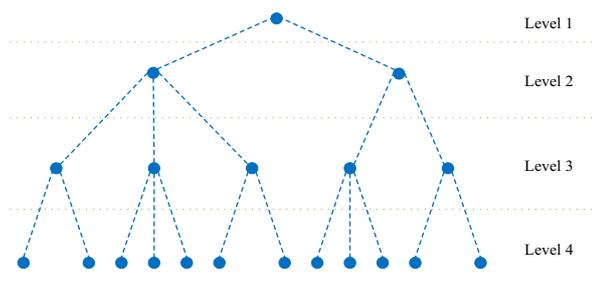}
\caption{The sketch map of the hidden tree structure. The hidden tree structure reflects the hidden order in the society, or the cascade controlling relationships among the individuals. Every node controls a few son nodes. $K$ is the average of the son nodes of all nodes except the leaf nodes and its value is 2.5.}\label{fig.hiddentree}
\end{figure}

Suppose that there are $N$ individuals in the society. Let $F_{i}$ represents the wealth of the i-th individual. All individuals are organized as a
tree. Assume that every individual except the leaf nodes has $K$ son nodes. If $K$ is a real number, for example, 2.5, then the ancestor node would certainly have 2 nodes and another one node with a probability of 0.5. Moreover, every individual has a ranking number, representing its level in the tree. Here the ranking number of root be $1$, denoted as $r(1) = 1$, the
ranking number of the first son-node of root would be 2, and so on. The hidden tree structure is illustrated in Fig. \ref{fig.hiddentree}.

We then randomly choose two individuals to simulate the transaction between them. For each transaction, there is only a shortest path in the tree structure. Moreover, for every
transaction, we set the total value of one transaction as $1$, and the value would be assigned among the individuals who participate
the transaction. The assignment function would be relative to the ranking numbers of all participating
individuals. For simplicity, assume that the assignment function follows equation \ref{eq.value}.

\begin{equation}
g(r)=r^t
    \label{eq.value}
\end{equation}

Here $r$ is the ranking number and $t$ is the exponent.

Let the numbers of all the participating individuals be $l_{1}, l_{2},..., l_{p}$, then the assigned value of the $l_j$ node in one transaction would follow the equation \ref{eq.distributing}.

\begin{equation}
   F_{l_{j}}=\frac{{g(r(l_j ))}}{s}
\end{equation}\label{eq.distributing}

Here,
\begin{equation}
s = \sum\limits_p {g(r(l_p ))}
\end{equation}\label{eq.distributing2}
%
%

\begin{figure}[htbp]\begin{center}
\includegraphics[width=8cm,height=4cm]{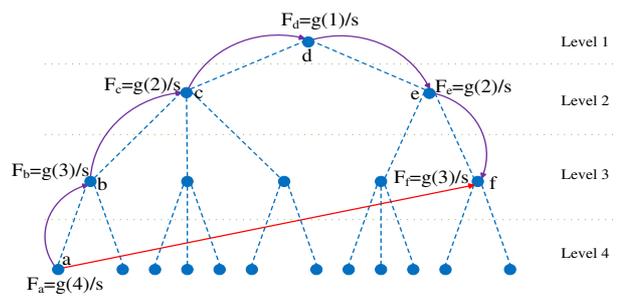}
\end{center}
\caption{The sketch map of a transaction on the hidden tree. The node $a$ proposes the transaction to node $f$. The red arrow indicates that node $a$ and node $f$ are chosen, the profit should be distributed among all the participants. The purple arrows indicate the shortest path and the participating nodes in the shortest path $b$, $c$, $d$, $e$ between $a$ and $f$, each of them will obtain profit $g(r)/s$. $s$ is the normalized factor and $s=g(4)+2g(3)+2g(2)+g(1)$. Every node would accumulate its wealth when the transactions continue to occur. }\label{fig.transactions}
\end{figure}

We use Fig. \ref{fig.transactions} to illustrate the rules of transactions.

Every individual would trade many times with the other individuals, so we use $A$ to represent the number of transactions that every node proposed.
The wealth of arbitrary node would be the summary of all the participated transactions.

Based on the model above, we prove that the distribution of wealth would obey the power law distribution by the simulation and theoretical analysis. That is,

\begin{equation}
p(F) \sim F^{-\gamma}
    \label{eq.gamma}
\end{equation}

\section{Experimental Results and Analysis}

There are four factors to the wealth distribution in this model. Firstly, the population of
individuals $N$; Secondly, the number of son nodes of every individuals $K$; Thirdly,  the times of
transactions for every individual $A$; Finally, the exponent value of the assignment function $t$.

\begin{figure}[!htbp]\begin{center}
\includegraphics[width=6.4cm,height=4.8cm]{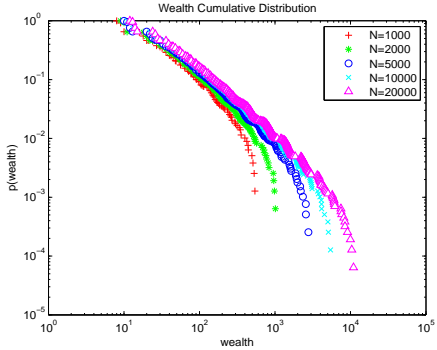}
\caption{\label{fig.pop}Wealth Cumulative Distribution When $N$ Varies. This figure uses the log-log coordination. All the curves approximate linear, indicating that the distribution of wealth is power-law. When the number of nodes $N$ varies, the exponent $\gamma$ does not vary, meaning that $N$ does not affect the exponent. At the tails of curves, there are an exponential cutoff. When $N$ increases, the location of the cutoff moves right. That is, the finity of nodes leads to the cutoff.}\end{center}
\end{figure}

For the first experiment, we set the population of society as $N=1000, 2000, 5000, 10000, 20000$, and the other parameters $K =
1.5$ , $A = 0.8$, $t=0$. We have such distributions as in Fig. \ref{fig.pop}.

From Fig. \ref{fig.pop}, we can see that no matter how the population varies, the distribution still follows the
power law.

\begin{figure}[!htbp]\begin{center}
\includegraphics[width=6.4cm,height=4.8cm]{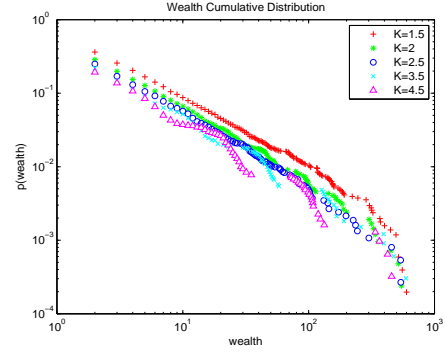}
\caption{\label{fig.sibling}Wealth Cumulative Distribution When $K$ Varies. From this figure, the curves approximate linear, indicating the power law distribution for different $K$. Some curves have waves which come from the fact that every node has integral son nodes, regardless of the real number $K$. }\end{center}
\end{figure}

The next experiment is designed to check the impact of the number of child nodes of every individual $K$. Here we set $K = 1.5, 2
, 2.5, 3.5, 4$, respectively, the other parameters $N = 10000, A = 0.4, g(r)=r^0=1$. We obtain the distributions as in Fig. \ref{fig.sibling}.

\begin{figure}[!htbp]\begin{center}
\includegraphics[width=6.4cm,height=4.8cm]{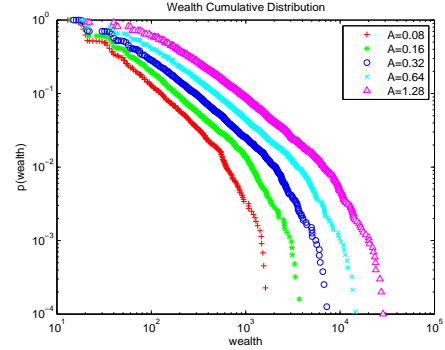}
\caption{\label{fig.activity}Wealth Distribution When $A$ Varies. For this figure, all the curves also approximate linear. When the number of transactions increases, the wealth of every node would increase with the same proportion, therefore, the curves can shift right when $A$ increase. Because the layer number is limited, the curves show a cut-off.}\end{center}
\end{figure}

The third experiment is designed to check the impact of the number of transactions $A$. Here we set $ A =
0.08 , 0.16, 0.32, 0.64, 1.28$, respectively, the other parameters $N = 10000, K = 1.5, t=0$. We obtain
such distributions as in Fig. \ref{fig.activity}.

The last experiment is designed to check the impact of the assignment function. We set
the parameter $t = -1, -0.5, 0, 0.5, 1$ respectively, and the other parameters  $N = 10000, K=2, A = 0.4$.

\begin{figure}[!htbp]\begin{center}
\includegraphics[width=6.4cm,height=4.8cm]{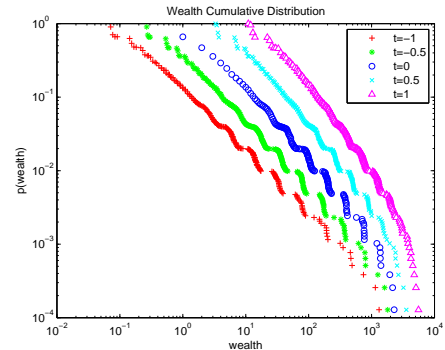}
\caption{\label{fig.t}Wealth Distribution When $t$ Varies. For different $t$, all the curves approximate linear, but the slopes are obviously different, suggesting that the assignment function can change the exponent value drastically.}\end{center}
\end{figure}

%
%
%

From these experiments above, we can see that the wealth distributions satisfy power law when the parameters
varies. These results imply that the wealth distributions are robust when the individuals are organized as a
tree.

\section{Theoretical Analysis}

Assume that the hidden tree is a complete K-tree, the number of nodes $N$ will be determined by the number of layers $M$, so we have,

\begin{equation}\label{eq.NandM}
   N{\rm{ = }}\frac{{{\rm{K}}^{{\rm{M + 1}}} {\rm{ - 1}}}}{{{\rm{K - 1}}}}
\end{equation}

To simplify the theoretical analysis, we consider a special example, i.e., assuming that only the leaf nodes can be chosen randomly, although the simulations allow all the nodes to be selected randomly. For any node, its ranking value $r$ determines the number of its offspring nodes.
Since the hidden tree is regular and complete, the nodes in every layer would
have the same probability to be selected to participate the transactions. Let us consider the root node in a 3-tree, assume that the root node has three branches $a$, $b$ and $c$,  we select two nodes in the leaf layer to perform one transaction. Assume that the first node belongs to the $a$ branch, thus when the second node also belongs to the $a$ branch, the root would not participate the transaction; when the second node belongs to the other branches, the root would participate. Therefore, the root node has a probability of $2/3$ to participate the transaction.
Similarly, the probability of nodes in r-th layer would be $(2/3)^{r}$. Because the number of nodes in
r-th layer would be $3^{r-1}$, according to the exponential combination method\cite{27,92,93,94}, the curve of the probability distribution of the participating transactions would be linear in the loglog coordination.

Generally, the probability of root node of a K-tree is $(K-1)/K$, and the probability of nodes in r-th layer
would be $((K-1)/K)^{r}$, and the number of nodes in the r-th layer would be $K^{r-1}$, the curve of the probability distribution of the participating transactions still are linear in loglog coordination.

Mathematically, the probability of the participating transactions of a node in the r-th layer is,

\begin{equation}\label{eq.Dandr}
D(r) \sim {(\frac{K-1}{K})}^r
\end{equation}

For any node, the probability that it has a ranking number $r$ can be calculated by equation \ref{eq.m.exponetialdistribution}.

\begin{equation}\label{eq.m.exponetialdistribution}
    p(r)\sim \frac{K^{r-1}}{N} \ \ \ \     (1\leq r \leq M)
\end{equation}

So we have,

\begin{equation}\label{eq.k.m.distribution2}
p(D) \sim D^{-1 - \frac{ln(K)}{ln(K)- ln(K-1)}}
\end{equation}

So we have, the exponent $\gamma = 1 + \frac{ln(K)}{ln(K)- ln(K-1)}$. When $K=2$, $\gamma=2$. According to equation \label{eq.k.m.distribution2}, the slight changes of $K$ does not change $\gamma$ greatly.

Notice the equation \ref{eq.value}, when $t=0$, we have,
\begin{equation}\label{eq.pf.f}
p(F) \sim  F^{-1 - \frac{ln(K)}{ln(K)- ln(K-1)}}
\end{equation}

That is, the wealth distribution would obey power law.

Notice that we can create a scale-free network by adding a simple rule that the first node of every transaction would link to all the participants. Obviously, the degree distribution would be the same as the distribution of the variant $D$. Moreover, the average shortest paths would vary according to the parameter $A$\cite{171,151}.

In general, the proposed model can produce the scale-free network, simultaneously, the power-law wealth distribution.

\section{Discussions and Conclusions}

From the theoretical analysis and the simulation experiments, we can draw a conclusion that the hidden tree structure does produce the power-law wealth distribution. Compared to the positive-feedback-like models, this model has some advantages. Firstly, this model has total parallelism. The actions of every nodes are not dependent to the others', so that the growth is not yet a necessary assumption. Secondly, the investment is not the only source to determine the wealth distribution. That is, the final individual wealth does not certainly come from the advantages of initial wealth. Creativity, i.e., creating a new branch of the hidden tree, can also produce immense wealth. Thirdly, the wealth distribution does not rely on the unstable positive feedback. By contraries, this model allows the stable structure, the hidden tree. Finally, the model is compatible to the Matthew effect. The process of generating a branch can be regarded as the examples of the Matthew effect.

Moreover, this hidden tree model means a lot to the ideas on the wealth inequality. From this model, we can see that the social structure leads to the wealth inequality. That is, according to the inverse theorem, if we want to achieve the result of wealth evenness, we have to destroy all the social/economic structures. Actually, the social/economic structure relates to the labor divisions. Since the labor divisions can empower the labor efficiency hundreds of thousands of times, once the labor divisions were destroyed, the products would drop drastically, that means, the society would experience a serious economic depression. This conclusion can be supported by the early civilization. Thousands of years ago, human beings accepted the civilizations even if it often means the cruel slavery, only because the social organization can provide more products, i.e., more surviving opportunities to all. Since our ancestors abandoned the wealth evenness, or say, zero wealth, we can not go back.

When the exponent of the assignment function varies, the exponent of power law of wealth distribution also varies. That is, the assignment function can not foundationally change the power-law wealth distribution, however, it still can adjust the inequality to some extent by increasing more assigned value to the low-level nodes.

In conclusion, although this paper drawn a pessimistic conclusion, however, it simultaneously proved that human beings are unnecessary to defy the natural principles, because we can modify the wealth distributing processes, by taxing or other policies, so that the products increase and are shared. Although the economic inequality is inevitable, even some regard it vicious, it still contributes much to our world.

\acknowledgments
Thanks to Dr. Oskar Burger, Dr. Chunlai Zhou, Dr. Baobin Wang  and Dr. Zhongxun Zhu for valuable discussions and to Dr. Ting Hu for English improvement.
Supported by
the State Key Laboratory of  Networking and Switching Technology (No. SKLNST-2010-1-04) and the National Natural Science Foundation of China (No.60803095).

\bibliographystyle{eplbib}
\bibliography{ComplexNetwork}

\end{document}